# Darboux-integration of $i\dot\rho = [H, f(\rho)]$

Nikolai V. Ustinov[1,a], Sergiej B. Leble[1,2,b], Marek Czachor[2,3,c], and Maciej Kuna[2,d]

[1] *Theoretical Physics Department, Kaliningrad State University, Kaliningrad, Russia*
[2] *Wydział Fizyki Technicznej i Matematyki Stosowanej*
*Politechnika Gdańska, ul. Narutowicza 11/12, 80-952 Gdańsk, Poland*
[3] *Arnold Sommerfeld Institut für Mathematische Physik*
*Technische Universität Clausthal, 38678 Clausthal-Zellerfeld, Germany*
[a] e-mail: n_ustinov@mail.ru
[b] e-mail: leble@mifgate.pg.gda.pl
[c] e-mail: mczachor@pg.gda.pl
[d] e-mail: maciek@mifgate.pg.gda.pl

A Darboux-type method of solving the nonlinear von Neumann equation $i\dot\rho = [H, f(\rho)]$ is developed. An explicit construction shows that self-scattering solutions are a generic property of such nonlinearities. An infinite-dimensional solution is constructed.

PACS: 05.45.-a, 05.70.Ln, 03.65.Bz

## I. INTRODUCTION

The equation

$$i\dot X = [Y, f(X)], \qquad (1)$$

plays an important role in different branches of quantum and classical physics. First of all, if $f$ satisfies $f(0)=0$, $f(1)=1$, and $X = X^\dagger = X^2$ then $f(X) = X$ [1] and the equation is simply the linear von Neumann equation for pure states. If $X = |\psi\rangle\langle\psi|$ and $Y = H$ is a Hamiltonian, Eq. (1) is physically equivalent to the Schrödinger equation

$$i|\dot\psi\rangle = H|\psi\rangle \qquad (2)$$

for a large class of *nonlinear* functions $f$.

Second, Eq. (1) can be written for a large class of functions $f(X)$ as the Heisenberg equation of motion

$$i\dot X = [X, h(X)], \qquad (3)$$

with Hamiltonian $h(X)$. (Assuming, for example, $f(X) = -X^n$ gives $h(X) = X^{n-1}Y + X^{n-2}YX + ... + YX^{n-1}$.) The equations of this form are often used in nonlinear quantum optics.

Third, it is known that the linear von Neumann equation can be considered as a Lie-Poisson Hamiltonian system. In the case of classical "extensive" statistics, the Hamiltonian function is given by the average energy $H_1 = \text{Tr}(H\rho)$. Yet, there are statistical problems that are naturally described by *nonextensive* statistics in terms of the Tsallis $q \neq 1$ entropies and "average" energies $H_q = \text{Tr}(H\rho^q)/\text{Tr}\,\rho^q$ [2]. One of the remarkable properties of $q$-statistics is the $q$-independence of standard geometric structures associated with equilibrium thermodynamics. Extending this observation from equilibria to non-equilibria, one finds that $H_q$ is a Hamiltonian function for a Lie-Poisson dynamics [3]:

$$i\dot\rho = [H, f(\rho)], \qquad (4)$$

where $f(\rho) = C_q(\rho)\rho^q$ and $C_q(\rho)$ is a Casimir function satisfying $C_1(\rho) = 1$. The von Neumann equation (4) with arbitrary nonlinearity is Lie-Poissonian with the Hamiltonian function $H_f = \text{Tr}\big(Hf(\rho)\big)$.

Fourth, the "Euler equations"

$$i\dot X = [Y, X^2], \qquad (5)$$

appear in several contexts. The best known physical example (with $Y_{kl} = \delta_{kl}/I_l$, $X_{mn} = i\varepsilon_{klm}J_k$) is given by the Euler equations for a freely rotating rigid body. Less known and often more abstract versions of the Euler equations are related to the Lie-Poisson equations occuring in fluid dynamics [4,5], the Nahm equations in non-abelian gauge theories [6], and the $N$–wave equations for electromagnetic waves in nonlinear media [7]. Quite recently the equation of the form

$$i\dot X = [Y, X^3] \qquad (6)$$

was discovered in the context of symmetries of the Euler equation [8].

General equations of the form (4) appeared in the context of nonlinear Nambu-type theories [9]. Nonlinear Lie-Poisson density matrix equations were applied in quantum mechanics with mean-field backgrounds [10] and nonlinear quantum mechanics [10,11]. Solutions of these equations were used as models of non-completely-positive nonlinear maps which, nevertheless, satisfied physical conditions widely believed to be equivalent to complete positivity [12]. Finally, the Lie-Poisson density matrix techniques of extending nonlinear evolutions of subsystems to entangled states proved to have applications in quantum information theory [13].

Although the literature devoted to the Euler equation is quite extensive [4,5,14], analytical methods were only recently applied to its density matrix version [15–17,20]. The constraints imposed on density matrices ($\rho^\dagger = \rho$, $\rho \geq 0$, $\text{Tr}\,\rho = 1$) and Hamiltonians ($H^\dagger = H$, $H \geq 0$, unboundedness) require techniques which are not based on standard integration via quadratures and the similar, since the systems in question are generically infinite-dimensional. The technique used in [15–17,20] is an appropriate modification of the dressing method [22,23] and, in particular, of the version based on a binary Darboux transformation [24–26].

The Darboux-type method of integration of the Euler-von Neumann equation

$$i\dot\rho = [H, \rho^2] \qquad (7)$$

introduced in [15] led to the discovery of the so-called self-scattering solutions. The process of self-scattering continuously interpolates between two asymptotically linear evolutions. This very characteristic property was found in all the nontrivial solutions of Eq.(7) obtained by the dressing method.

A problem that remained open in all the previous works was how to obtain solutions of Eq.(4) with other values of the Tsallis parameter $q$. In fact, the case $q = 2$ was not very interesting from the point of view of nonextensive applications since the parameters involved in analysis of concrete physical situations were either close to 1 or $0 < q < 1$. Of special interest turned out to be the value $q = 1/2$ due to its plasma physics applications. Below we present an extension of the Darboux technique to the nonlinear von Neumann equation for *all* functions $f$ for which $f(\rho)$ is well defined [1].

## II. LAX PAIRS AND THEIR DARBOUX COVARIANCE

We begin with the overdetermined linear system (Lax pair)

$$z_\lambda \langle\psi| = \langle\psi|(\rho - \lambda H) \qquad (8)$$
$$-i\langle\dot\psi| = \frac{1}{\lambda}\langle\psi|A \qquad (9)$$

where $A$, $\rho$, $H$ are operators, $\lambda$, $z_\lambda$ are complex numbers, and $\langle\psi|$ is a "bra" associated with an an element of a Hilbert space. The operators $\rho$ and $H$ will typically play the roles of density matrices and Hamiltonians, respectively, but one can also think of them as just some operators without any particular quantum mechanical connotations. The compatibility condition of the Lax pair is

$$i\dot\rho = [H, A], \qquad (10)$$
$$[A, \rho] = 0. \qquad (11)$$

It is convenient to introduce two additional Lax pairs

$$z_\mu|\varphi\rangle = (\rho - \mu H)|\varphi\rangle, \qquad (12)$$
$$i|\dot\varphi\rangle = \frac{1}{\mu}A|\varphi\rangle, \qquad (13)$$

$$z_\nu\langle\chi| = \langle\chi|(\rho - \nu H), \qquad (14)$$
$$-i\langle\dot\chi| = \frac{1}{\nu}\langle\chi|A. \qquad (15)$$

The method of solving (4) is based on the following

**Theorem.** Assume $\langle\psi|$, $|\varphi\rangle$, and $\langle\chi|$ are solutions of (8), (9),(12)–(15) and $\langle\psi_1|$, $\rho_1$, $A_1$ are defined by

$$\langle\psi_1| = \langle\psi|\Big(\mathbf{1} - \frac{\nu - \mu}{\lambda - \mu}P\Big), \qquad (16)$$
$$\rho_1 = \Big(\mathbf{1} + \frac{\mu - \nu}{\nu}P\Big)\rho\Big(\mathbf{1} + \frac{\nu - \mu}{\mu}P\Big), \qquad (17)$$
$$A_1 = \Big(\mathbf{1} + \frac{\mu - \nu}{\nu}P\Big)A\Big(\mathbf{1} + \frac{\nu - \mu}{\mu}P\Big), \qquad (18)$$
$$P = \frac{|\varphi\rangle\langle\chi|}{\langle\chi|\varphi\rangle}. \qquad (19)$$

$$(20)$$

Then

$$z_\lambda\langle\psi_1| = \langle\psi_1|(\rho_1 - \lambda H), \qquad (21)$$
$$-i\langle\dot\psi_1| = \frac{1}{\lambda}\langle\psi_1|A_1. \qquad (22)$$

*Proof*: Substitution of Eq. (16) into Eq. (21) defines $\rho_1$ by

$$\rho_1 = \rho + (\mu - \nu)[P, H]. \qquad (23)$$

The proof that (23) and (17) are equivalent can be found in [16]. To prove Eq. (22) one first notices that the operator $P$ given by (19) satisfies the nonlinear equation

$$i\dot P = \frac{1}{\mu}(1 - P)AP - \frac{1}{\nu}PA(1 - P) \qquad (24)$$

and (22) follows from a straightforward calculation. $\square$

**Lemma.** If $A = f(\rho)$ [1], then

$$i\dot\rho_1 = [H, f(\rho_1)]. \qquad (25)$$

*Proof*: This statement follows from the compatibility condition for the Lax pair (21)–(22), if one takes into account

$$A_1 = \Big(\mathbf{1} + \frac{\mu - \nu}{\nu}P\Big)f(\rho)\Big(\mathbf{1} + \frac{\nu - \mu}{\mu}P\Big) = f(\rho_1). \quad \square$$

*Comment*: Having one solution $\rho$ one produces a new solution $\rho_1$ of the nonlinear von Neumann equation (4). The procedure can be further iterated, $\rho \to \rho_1 \to \rho_2 \to \ldots$, in a direct analogy to the Darboux method of generating multi-solitons [18] or supersymmetric quantum mechanics [19].

If $\rho$ is a density matrix and $\nu = \bar\mu$ one can put $\langle\chi| = \langle\varphi|$. In this case $\rho_1$ is also a density matrix and spectra of $\rho$ and $\rho_1$ are identical.

# III. SELF-SCATTERING SOLUTIONS FOR A GENERAL NONLINEARITY

We will now show that self-scattering solutions constructed in [15] for $f(\rho) = \rho^2$ are a generic property of the general nonlinear von Neumann equation (4).

One begins with a seed solution satisfying

$$f(\rho) - a\rho = \Delta_a \qquad (26)$$

where $[\Delta_a, H] = 0$ and $\Delta_a$ is not a multiple of the identity. The solution satisfies

$$i\dot\rho = [H, f(\rho)] = a[H, \rho] \qquad (27)$$

and

$$\rho(t) = e^{-iaHt}\rho(0)e^{iaHt}. \qquad (28)$$

Taking the Lax pair with $\nu = \bar\mu$ and repeating the construction from [15] one arrives at the self-scattering solution

$$\rho_1(t) = e^{-iaHt}\Big(\rho(0) + (\mu - \bar\mu)F_a(t)^{-1}e^{-\frac{i}{\bar\mu}\Delta_a t} \\ \times [|\varphi(0)\rangle\langle\varphi(0)|, H]e^{\frac{i}{\bar\mu}\Delta_a t}\Big)e^{iaHt} \qquad (29)$$

where

$$F_a(t) = \langle\varphi(0)|\exp\Big(i\frac{\mu - \bar\mu}{|\mu|^2}\Delta_a t\Big)|\varphi(0)\rangle \qquad (30)$$

and $|\varphi(0)\rangle$ is an initial condition for the Lax pair.

As an example consider the harmonic oscillator Hamiltonian $H = \omega \sum_{n=0}^{\infty} n|n\rangle\langle n|$ and

$$f(\rho) = \rho^q - 2\rho^{q-1} \qquad (31)$$

where $q \in \mathbf{R}$. To construct the operator $\Delta_a$ we first note that for any $q$ the equation

$$x^q - 2x^{q-1} - x + 2 = (x^{q-1} - 1)(x - 2) = 0 \qquad (32)$$

has two positive solutions, $x = 1$ and $x = 2$, which will be used in the construction of a seed $\rho$. Choosing $a = 1$ we define

$$\Delta_1 = f(\rho) - \rho = \rho^q - 2\rho^{q-1} - \rho. \qquad (33)$$

The next task is to choose a seed $\rho$ such that $[\rho, H] \neq 0$ and $[\Delta_1, H] = 0$. We do this as follows. Let $|0\rangle, |1\rangle, |2\rangle$, be the three lowest energy eigenstates of $H$ and take

$$\rho(0) = \frac{3}{2}\Big(|0\rangle\langle 0| + |2\rangle\langle 2|\Big) \\ + \frac{7}{4}|1\rangle\langle 1| \\ - \frac{1}{2}\Big(|2\rangle\langle 0| + |0\rangle\langle 2|\Big). \qquad (34)$$

We find $[\rho(0), H] \neq 0$ and

$$\Delta_1 = -2\Big(|0\rangle\langle 0| + |2\rangle\langle 2|\Big) \\ + \Big(-2 + \tfrac{1}{4}[1 - (\tfrac{4}{7})^{1-q}]\Big)|1\rangle\langle 1|. \qquad (35)$$

The third eigenvalue of $\rho(0)$ is $7/4$ and therefore the density matrix is not yet normalized. Now take $\mu = i\sqrt{3}/(4\omega)$. The eigenvalues of $\rho - \mu H$ are $5/4 + i\sqrt{3}/4$ and $7/4 + i\sqrt{3}/4$, the latter being twice degenerated. The initial condition for the Lax pair is taken to be a linear combination of the two orthonormal eigenvectors corresponding to $7/4 + i\sqrt{3}/4$:

$$|\varphi(0)\rangle = \frac{i + \sqrt{3}}{4}|0\rangle + \frac{1}{\sqrt{2}}|1\rangle + \frac{i - \sqrt{3}}{4}|2\rangle. \qquad (36)$$

As a general rule self-scattering solutions can occur only for $|\varphi(0)\rangle$ which are not eigenstates of $\Delta_a$. Here it means that $q \neq 1$ which is consistent with the fact that for $q = 1$ the von Neumann equation is linear.

Now we are in position to explicitly write the self-scattering solution for any $q$. We start with the seed solution

$$\rho(t) = e^{-iHt}\rho(0)e^{iHt} \qquad (37)$$

and obtain

$$\rho_1(t) = e^{-iHt}\rho_{\text{int}}(t)e^{iHt} \qquad (38)$$

with

$$\rho_{\text{int}}(t) = \sum_{m,n=0}^{2} r(t)_{mn}|m\rangle\langle n|, \qquad (39)$$

$$\begin{pmatrix} r_{00} & r_{01} & r_{02} \\ r_{10} & r_{11} & r_{12} \\ r_{20} & r_{21} & r_{22} \end{pmatrix} = \begin{pmatrix} 3/2 & -\xi(t) & \zeta(t) \\ -\bar\xi(t) & 7/4 & \xi(t) \\ \bar\zeta(t) & \bar\xi(t) & 3/2 \end{pmatrix}, \qquad (40)$$

$$\xi(t) = \frac{(-3i + \sqrt{3})e^{\omega_q t}}{4\sqrt{2}(1 + e^{2\omega_q t})}, \qquad (41)$$

$$\zeta(t) = \frac{1 - i\sqrt{3} - 2e^{2\omega_q t}}{4(1 + e^{2\omega_q t})}, \qquad (42)$$

$\omega_q = [(4/7)^{1-q} - 1]\omega/\sqrt{3}$, $\omega_q \geq 0$ for $q \geq 1$, $\omega_q \leq 0$ for $q \leq 1$. The self-scattering asymptotics is

$$\xi_{q \neq 1}(\pm\infty) = 0, \qquad (43)$$
$$\zeta_{q>1}(+\infty) = \zeta_{q<1}(-\infty) = -1/2, \qquad (44)$$
$$\zeta_{q>1}(-\infty) = \zeta_{q<1}(+\infty) = (1 - i\sqrt{3})/4. \qquad (45)$$

Let us note that the seed solution $\rho(t)$ we have started with reappers in the asymptotic states

$$\rho_1 \to \rho, \quad \text{for } q > 1, \; t \to +\infty \qquad (46)$$
$$\rho_1 \to \rho, \quad \text{for } q < 1, \; t \to -\infty. \qquad (47)$$

The initial condition at $t = 0$ is the same for all $q$.

It is interesting that, somewhat counter-intuitively, equations with very different $q$ may lead to evolutions which are practically indistinguishable. Indeed, for $q \to +\infty$ we get $\omega_q \to \infty$ and therefore the transition around $t = 0$ between the asymptotic linear evolutions takes the less time the greater $q$; for $q \to -\infty$ we get $\omega_q \to -\omega/\sqrt{3}$. By the same method but with a different choice of $|\varphi(0)\rangle$ one can generate solutions whose self-scattering takes place in a neighborhood of an arbitrarily chosen $t = t_0$ (for details cf. the discussion of $f(\rho) = \rho^2$ given in [20]).

Time scales involved in self-scattering are best illustrated by average position of the harmonic oscillator as a function of time. Fig. 1 shows $\langle x \rangle = \mathrm{Tr}\,(\hat{x}\rho_1)/\mathrm{Tr}\,\rho_1$, $\hat{x} = (a + a^\dagger)/\sqrt{2}$ for different values of $q$. In Fig. 2 the self-scattering is explicitly seen in the contour plot of a Harzian [21], a 3D surface representing the self-scattering time dependent probability density in position space as a function of time.

## IV. INFINITE-DIMENSIONAL EXAMPLE

In this section we will show that the above technique is not limited to matrix cases. The example is, perhaps, rather artificial but at least clearly shows that a strategy leading to self-scattering solutions involving infinite-dimensional subspaces is possible. We are not interested in the trivial situation where the dynamics is "reducible" i.e. can be decomposed into a direct sum of finite-dimensional evolutions. This could be done immediately on the basis of the analysis given in the previous section.

Assume that spectrum of the Hamiltonian $H$ contains a discrete part $\{E_n\}_{n=1}^\infty$. One technical assumption we will need [27] is a symmetry of the spectrum of $H$, at least when restricted to the subspace we are interested in. By this we mean that the spectral representation of $H$ is

$$H = \sum_{n=1}^\infty E_n \big(|n, +\rangle\langle n, +| - |n, -\rangle\langle n, -|\big) + \ldots \quad (48)$$

where the dots stand for the remaining part of the spectrum.

Equation $i\dot{\rho} = [H, f(\rho)]$ is Lie-Poissonian with the Hamiltonian function $H_f = \mathrm{Tr}\,Hf(\rho)$. In nonextensive statistics [2,3] one considers such generalized averages where $f(\rho) = \rho^q$. From the nonextensive point of view it is essential that it is $\mathrm{Tr}\,H\rho^q$, $\mathrm{Tr}\,\rho^q = 1$, and not $\mathrm{Tr}\,H\rho$ which plays the role of the energy.

As a consequence, it is natural to treat $\varrho = f(\rho)$ normalized by $\mathrm{Tr}\,f(\rho) = 1$ as a physical density matrix associated with the solution $\rho$ since then the Hamiltonian function has an interpretation in terms of average energy. In what follows we shall assume that $f(-x) = f(x)$. For Tsallis-type description one can take $f(x) = x^q$, $q = 2n/(2n \pm 1)$ and the linear limit is $n \to \infty$.

With the above assumptions we can take the seed solution $\rho$ in the form of an infinite matrix

$$\rho = \begin{pmatrix} a_1 & 0 & 0 & 0 & 0 & 0 & \ldots \\ 0 & -a_1 & 0 & 0 & 0 & 0 & \ldots \\ 0 & 0 & a_2 & 0 & 0 & 0 & \ldots \\ 0 & 0 & 0 & -a_2 & 0 & 0 & \ldots \\ 0 & 0 & 0 & 0 & a_3 & 0 & \ldots \\ 0 & 0 & 0 & 0 & 0 & -a_3 & \ldots \\ \vdots & \vdots & \vdots & \vdots & \vdots & \vdots & \ddots \end{pmatrix} \quad (49)$$

and the Hamiltonian

$$H = \begin{pmatrix} b_1 & c_1 & 0 & 0 & 0 & 0 & \ldots \\ \bar{c}_1 & -b_1 & 0 & 0 & 0 & 0 & \ldots \\ 0 & 0 & b_2 & c_2 & 0 & 0 & \ldots \\ 0 & 0 & \bar{c}_2 & -b_2 & 0 & 0 & \ldots \\ 0 & 0 & 0 & 0 & b_3 & c_3 & \ldots \\ 0 & 0 & 0 & 0 & \bar{c}_3 & -b_3 & \ldots \\ \vdots & \vdots & \vdots & \vdots & \vdots & \vdots & \ddots \end{pmatrix}. \quad (50)$$

Any Hamiltonian whose eigenvalues are $E_k = \pm\sqrt{b_k^2 + |c_k|^2}$ can be written in this way in some basis. Let us note that for $c_k \neq 0$ one finds $[\rho, H] \neq 0$ but nevertheless $[\varrho, H] = [f(\rho), H] = 0$ which means that the seed solution $\rho$ is stationary. The next task is to choose the parameters $a_k$ and $\nu$ in a way guaranteeing that the projector

$$P = \frac{|\chi_\nu\rangle\langle\chi_\nu|}{\langle\chi_\nu|\chi_\nu\rangle} \quad (51)$$

where

$$z_\nu\langle\chi_\nu| = \langle\chi_\nu|(\rho - \nu H) \quad (52)$$

will have nonzero matrix elements between any two eigenvectors of $H$. By construction the same will hold for

$$\rho[1] = \rho + (\bar{\nu} - \nu)[P, H] \quad (53)$$

and the nonlinear evolution will be "infinite-dimensional and irreducible" i.e. will involve transitions between all the eigenvectors of $H$ which span the infinite-dimensional subspace.

Take two constants $\alpha$, $\beta$ satisfying $b_k = \alpha a_k$ and $|c_k|^2 = \beta^2 a_k^2$. The eigenvalues of $H$ are $E_k^\pm = \pm a_k\sqrt{\alpha^2 + \beta^2}$. The normalization of $\varrho$ means $\sum_{n=1}^\infty f(E_k^+/\sqrt{\alpha^2 + \beta^2}) = 1/2$. It turns out that with the above choice of $\rho$ the condition of "infinite dimensionality and irreducibility" can be satisfied only if $z_\nu = 0$. We look for $\nu$ satisfying

$$(\alpha^2 + \beta^2)\nu^2 - 2\alpha\nu + 1 = 0. \quad (54)$$

and find $\nu_\pm = \frac{\alpha \pm i\beta}{\alpha^2 + \beta^2}$.

The eigenvector $|\chi_\nu\rangle$ corresponding to $z_\nu = 0$ is of the form

$$|\chi_\nu\rangle = \begin{pmatrix} u_1 w \\ u_2 w \\ \vdots \\ u_k w \\ \vdots \end{pmatrix} \qquad (55)$$

where

$$w = \frac{1}{\sqrt{2}} \begin{pmatrix} 1 \\ -i \end{pmatrix}. \qquad (56)$$

We finally obtain

$$\rho[1] = \begin{pmatrix} \rho[1]_{11} & \rho[1]_{12} & \rho[1]_{13} & \cdots \\ \rho[1]_{21} & \rho[1]_{22} & \rho[1]_{23} & \cdots \\ \rho[1]_{31} & \rho[1]_{32} & \rho[1]_{33} & \cdots \\ \vdots & \vdots & \vdots & \ddots \end{pmatrix} \qquad (57)$$

where

$$\rho[1]_{kl} = a_k \delta_{kl} \begin{pmatrix} 1 & 0 \\ 0 & -1 \end{pmatrix} \qquad (58)$$
$$+ \frac{\beta F_{kl}}{G} \left[ \frac{a_k}{\alpha - i\beta} \begin{pmatrix} i & 1 \\ 1 & -i \end{pmatrix} + \frac{a_l}{\alpha + i\beta} \begin{pmatrix} -i & 1 \\ 1 & i \end{pmatrix} \right],$$

where

$$G = \sum_{n=1}^{\infty} |u_n|^2 e^{2\beta f(a_n) t}, \qquad (59)$$

$$F_{kl} = u_k \bar{u}_l e^{-i\alpha[f(a_k) - f(a_l)]t + \beta[f(a_k) + f(a_l)]t}. \qquad (60)$$

The solution is "infinite-dimensional and irreducible". It seems that this is the first example of such a nonlinear dynamics one can find in the literature. The eigenvalues of $\rho[1]$ are the same as those of $\rho$. The form of $G$ and $F_{kl}$ shows that $\rho[1]$ is again a self-scattering solution.

## V. COMMENTS

We have shown that: (1) General von Neumann equations with $f$-nonlinearities possess nontrivial solutions. (2) Independently of $f$ such solutions can be obtained by the dressing method. (3) The self-scattering solutions are qualitatively similar for different nonlinearities. (4) The nonlinear effects are well localized in time, transient and asymptotically the solutions correspond to those of linear von Neumann equations. (5) Even large modifications of nonlinearity may lead to small and very short-lived modifications of standard linear dynamics. (6) All nonlinearities $\sim \rho^q$ which are expected to be related to nonextesive statistics can be treated within the proposed formalism. (7) The formalism can be applied to genuinely infinite-dimensional systems.

In the light of these findings one may wonder whether it is possible to experimentally distinguish between a general $f$ and a *linear* $f$. Indeed, the fact that some experimental data are well fitted by a linear dynamics may only mean that a self-scattering has taken place in the past, or will take place in the future. If in addition the state is pure then its dynamics is given by the linear von Neumann (or Schrödinger) equation even in case the function $f$ is highly nonlinear. It follows that not only the results we have reported may prove useful as a technical tool in many branches of classical and quantum physics, but they also shed new light on the negative results of experiments searching for quantum nonlinearity.

In a separate paper we will describe a generalization of the above technique to a larger class of "nonabelian" nonlinearities [28].

Our work was supported by the Alexander von Humboldt Foundation (M.C.) and Nokia-Poland (N.V.U.).

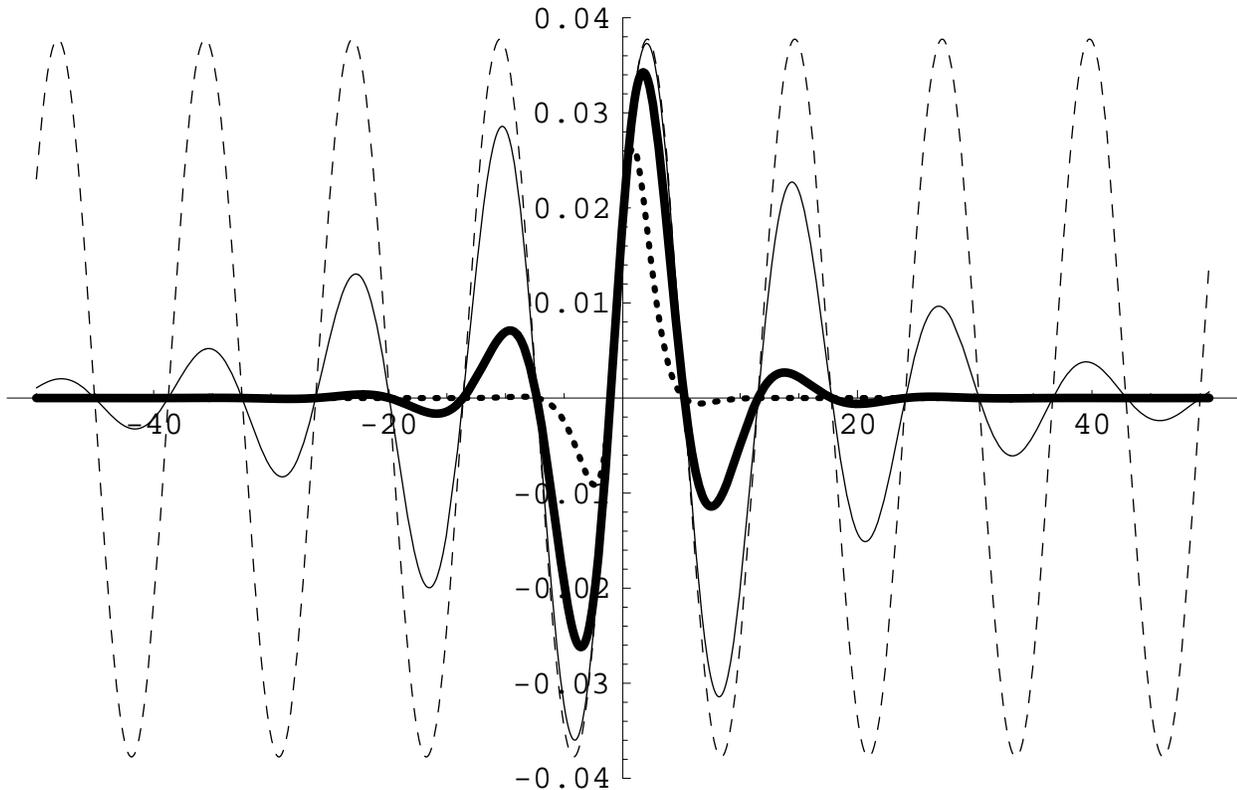

FIG. 1. $\langle x \rangle$ for $q = 1$ (dashed), $q = 2^{1/2}$ (thin solid), $q = \pi$ (dotted), and $q = -2$ (thick solid). All evolutions for $q > 1$ ($q < 1$) have identical asymptotic states and the same initial condition (all curves intersect at $t = 0$). The solution for $q = 1$ satisfies the same linear equation as the asymptotic states for $q \neq 1$.

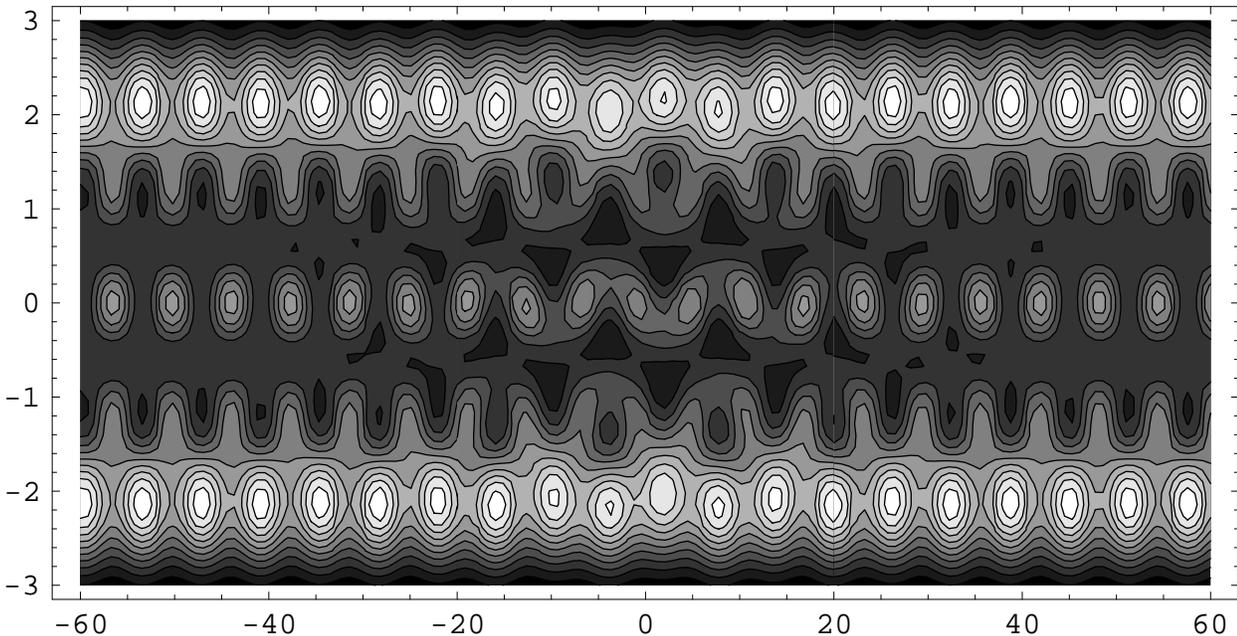

FIG. 2. Contour plot of a 3D surface representing probability density in position space $\langle x|\rho_1|x\rangle$ as a function of time for $q = 1/2$, $\omega = 1/2$, and $-60 < t < 60$. Continuous transition (self-scattering) between two solutions of the linear equation is clearly seen. $\rho_1$ has nonvanishing matrix elements in the subspace spanned by $|2\rangle$, $|3\rangle$, and $|4\rangle$.

7